\documentclass[a4paper]{article}

\usepackage{INTERSPEECH2022}
\usepackage{multirow}
\usepackage{makecell}
\usepackage{siunitx}
\usepackage{caption}
\newcolumntype{?}{!{\vrule width 1pt}}

\usepackage{comment}

\title{Mix and Match: An Empirical Study on Training Corpus Composition for Polyglot Text-To-Speech (TTS)}
\name{Ziyao Zhang$^1$, Alessio Falai$^1$, Ariadna Sanchez$^1$,  Orazio Angelini$^1$, Kayoko Yanagisawa$^1$}
\address{
  $^1$Alexa AI, Amazon
}
\email{\{zhaziyao,falai,cariadn,aorazio,yakayoko\}@amazon.com}

\begin{document}

\maketitle
\begin{abstract}
Training multilingual Neural Text-To-Speech (NTTS) models using only monolingual corpora has emerged as a popular way for building voice cloning based Polyglot NTTS systems. In order to train these models, it is essential to understand how the composition of the training corpora affects the quality of multilingual speech synthesis. In this context, it is common to hear questions such as “Would including more Spanish data help my Italian synthesis, given the closeness of both languages?”. Unfortunately, we found existing literature on the topic lacking in completeness in this regard. In the present work, we conduct an extensive ablation study aimed at understanding how various factors of the training corpora, such as language family affiliation, gender composition, and the number of speakers, contribute to the quality of Polyglot synthesis. Our findings include the observation that female speaker data are preferred in most scenarios, and that it is not always beneficial to have more speakers from the target language variant in the training corpus. The findings herein are informative for the process of data procurement and corpora building. 
\end{abstract}

\section{Introduction}
\label{sec:introduction}
Attempts to go beyond building single-speaker, monolingual Text-To-Speech (TTS) systems date far back in time \cite{javkin1989multilingual}. The advancements in Neural TTS (NTTS)  have seen the emergence of TTS models which are able to deliver high-quality multilingual and multi-speaker speech synthesis \cite{zhang2019learning,yu2016learning,do2021systematic}. These multilingual TTS models differ with each other significantly in terms of the level of flexibility of speaker and language variant (\textbf{LV} hereafter, which is defined as local variants of a language) combinations that they are able to deliver. 
In this paper, we define a multi-speaker multilingual TTS system, which enables any speaker in the training corpus to speak any LV present in the same corpus, as Polyglot TTS.

Multilingual TTS systems can generally be categorised into two realms, depending on whether cross-lingual voice cloning \cite{zhang2019learning}, defined as converting a certain speaker's voice into speaking a new language, is attempted. 
In this paper, we focus on the paradigm where multilingual TTS is realised via voice cloning.
In such a scenario, model training can be very challenging, because of the requirement of speaker and language disentanglement \cite{fong11improving}. Although it has been shown \cite{zhao2020towards} that having ``polyglot'' training corpora, which include recordings of the same speaker in multiple languages, improves multilingual synthesis, it is rarely practical to procure such data and the approach is unscalable. In light of this, the most realistic solution for building a voice cloning-based multilingual NTTS system is to train models using training corpora consisting of multiple monolingual speakers, which is the scenario we consider in this paper. 

There have been multiple studies on various aspects of multilingual NTTS model training using monolingual corpora. For example, \cite{zhang2019learning,nachmani2019unsupervised,himawan2020speaker,yang2022cross} introduce several representative NTTS architectures for building voice cloning-based TTS systems. To enable cross-lingual voice conversion, authors in \cite{williams2021exploring,wang2021accent} focus on achieving better disentanglement of several elements of speech, such as speaker characteristic and linguistic information. On addressing the data scarcity problems in some LVs, \cite{maniati2021cross,demirsahinunified} build multilingual TTS models for delivering TTS services in low-resource languages. 
Amongst these studies, one important topic that is largely missing is how the language and speaker composition of the training corpora affects the quality of multilingual speech from speakers present in the dataset. This paper's contribution fills the gap with large scale experiments comparing various training data compositions in terms of gender, language family mixture of speakers and the number of speakers. Our aim is to understand how these factors affect the target speaker's ability to speak the target language. We believe these results are valuable for constructing training corpora used for training Polyglot NTTS models. 

The rest of the paper is organised as follows. Section~\ref{sec:polyglot_tts} describes the architecture of the NTTS model we use for all our experiments. Section~\ref{sec:exp_design} details the experiment designs and the aim of each individual experiment. Section~\ref{sec:result_analysis} presents experiment results accompanied by analyses. Finally, Section~\ref{sec:conclusion} concludes the paper. 

\begin{comment}
\begin{table*}[h]
\caption{
\label{tab:locale_code}
Language variants and their corresponding codes.  
}
\begin{minipage}{\textwidth}
\centering
\begin{tabular}{l|l?l|l?l|l?l|l?l|l}
\toprule

\makecell{\textbf{Language} \\ \textbf{Variant}}& \makecell{\textbf{Code}} & \makecell{\textbf{Language} \\ \textbf{Variant}}  & \makecell{\textbf{Code}}   & \makecell{\textbf{Language} \\ \textbf{Variant}}  & \makecell{\textbf{Code}}    & \makecell{\textbf{Language} \\ \textbf{Variant}}  & \makecell{\textbf{Code}} &  \makecell{\textbf{Language} \\ \textbf{Variant}}  & \makecell{\textbf{Code}}
\\ \midrule

\makecell{Arabic\footnote{Our recordings are in Modern Standard Arabic (MSA).}}& \makecell{arb} & \makecell{Polish}  & \makecell{pl-PL}   & \makecell{Romanian}& \makecell{ro-RO} &  \makecell{European French} & \makecell{fr-FR} & \makecell{British English}  & \makecell{en-GB}    
\\ \hline

\makecell{Welsh}& \makecell{cy-GB} & \makecell{Korean}  & \makecell{ko-KR}   & \makecell{Italian}& \makecell{it-IT} & \makecell{Canadian French}  & \makecell{fr-CA} & \makecell{American English}  & \makecell{en-US}    
\\ \hline

\makecell{Danish}& \makecell{da-DA} & \makecell{Icelandic}  & \makecell{is-IS} 
& \makecell{German}& \makecell{de-DE} & \makecell{Norwegian}  & \makecell{nb-NO} & \makecell{Castilian Spanish}  & \makecell{es-ES}    
\\ \hline

\makecell{Russian}  &  \makecell{ru-RU}  & \makecell{Swedish}  & \makecell{sv-SE}  & \makecell{Turkish} & \makecell{tr-TR} & & &  
\\ \bottomrule
\end{tabular}
\end{minipage}
\vspace{-.3cm}
\end{table*}
\end{comment}

\section{NTTS architecture for Polyglot TTS}
\label{sec:polyglot_tts}

\subsection{Model architecture}
\label{sec:architecture}
\begin{figure}[t]
  \centering
  \includegraphics[width=.8\linewidth]{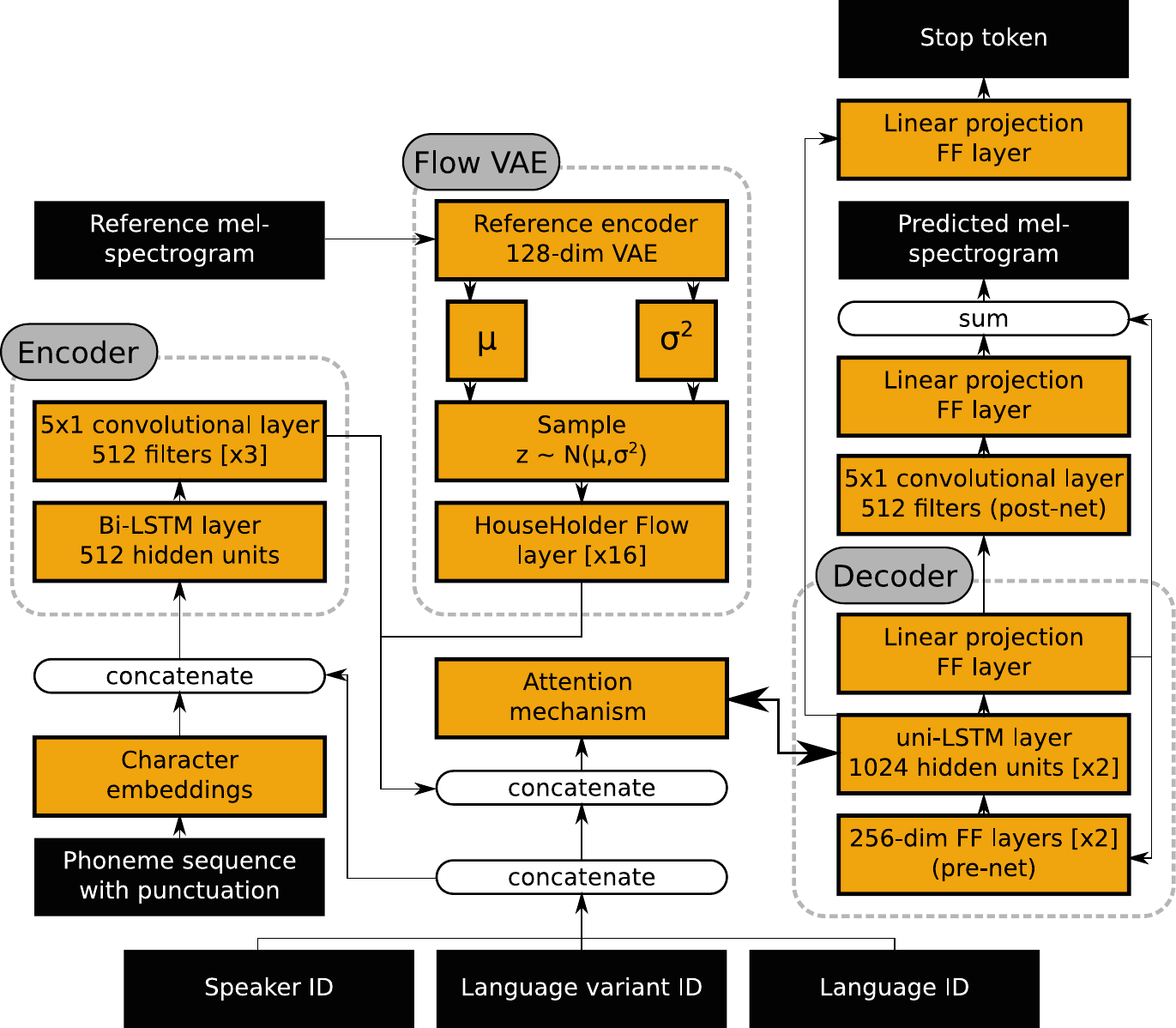}
  \caption{The model architecture used in all our experiments.}
  \label{fig:arch}
\vspace{-.2cm}
\end{figure}

Our model is based on Tacotron 2 \cite{tacotron2}, i.e., it is sequence-to-sequence and attention-based \cite{sequence-to-sequence}, and consists of $3$ main parts: an encoder, a decoder and an attention mechanism (Figure~\ref{fig:arch}).

The encoder takes as input a sequence of phonemes and embeds them into $256$-dimensional latent vectors. These phoneme embeddings are then passed through a stack of $3$ convolutional blocks, composed of 1D convolutions, ReLU activation, batch normalisation and dropout. Each 1D convolution in the encoder has $512$ filters and a kernel size of $5$, while the dropout rate for each block is set to $0.3$. After that, hidden representations go through a single bi-LSTM unit, having an output dimension of $512$. We employ a commonly used Tacotron 2 style decoder, similarly to ~\cite{merritt2018comprehensive,latorre2019effect}. The attention mechanism feeds into the LSTM decoder. The output is passed through two linear projections. One projection leads to the stop token prediction. The other projection feeds into both the 512 CNN filters of the post-net and the 256-d feed-forward pre-net, which in turn feeds into the LSTM layer of the decoder.
As for the attention mechanism, we rely on the default location-sensitive cell \cite{location-sensitive-attention}, with the same settings reported in \cite{latorre2019effect}, and we accumulate the attention weights at each decoder time-step to speed up the training procedure.

In addition, we employ specialized lookup tables to enable multi-speaker cross-lingual synthesis. In particular, we allocate $256$-dimensional embeddings to condition the encoder on speaker, language and LV information, by concatenating them at each time-step.  Finally, a custom reference encoder \cite{reference-encoder} is used to model speaking styles, where we consider “style” as anything that cannot be inferred solely from text. The reference encoder is composed of a VAE \cite{vae} that takes as input an 80-d vector representing a mel-spectrogram frame and outputs a 128-d latent vector. The VAE is then followed by 16 \textit{householder} flow layers \cite{householder-flow}.
The reference encoder’s output is then concatenated to the encoder’s hidden states. Finally, we use a universal vocoder based on \cite{jiao2021universal}.

\subsection{Data preparation and training strategy}
Our Polyglot model takes as input a sequence of phonemes that are generated by passing input text through a TTS frontend. Each LV has its own phone set with punctuation, although all LVs share the same service tokens, such as word boundary and sentence start tokens. The training targets are mel-scaled spectrograms extracted from audio recordings. 

We use a common set of training strategies, such as the optimiser, learning rate, gradient clipping, training loss, etc., for all models trained in this paper. As this is not the focus of our study, we refer readers to ~\cite{merritt2018comprehensive} for more details on model training settings.

\section{Experiment and evaluation designs}
\label{sec:exp_design}

\subsection{Terminology}
To facilitate the descriptions of experiment designs, we define the following terms and acronyms. 

\begin{itemize}
\itemsep0em 
\item Target Speaker (\textbf{TS}):  The speaker whose voice is used for synthesising in multiple LVs.
\item Target Language variant (\textbf{TL}): The LV(s) which the Polyglot model is intended to speak.
\item Speaker of Target Language variant (\textbf{STL}):  Speaker(s) whose data is included in Polyglot model training, and their native LV is a target LV. 
\item Assisting Speaker (\textbf{AS}): Speaker(s) whose data is included in model training, but their native LV is not one of the target LVs. 
\end{itemize}

Note that we use the ISO 639 standard for denoting codes of language variants, i.e., they are noted in the format of \{language code\}\_\{LV CODE\}, except for Arabic (arb).

\subsection{Core dataset}
\label{sec:baseline_dataset}
Our experiments found that in order to ensure that audio samples are free from conspicuous quality artefacts such as distorted speaker identity, which could bias the evaluation of overall quality of Polyglot synthesis, a core subset of the available training corpus is needed. Note that finding an optimised strategy for maintaining consistent speaker identity during voice cloning is beyond the scope of this paper. We use the same core data composition, selected empirically, for all experiments conducted in this study. In particular, the core dataset we use includes 40 speakers whose native languages span 11 LVs across 3 languages (English, French, Spanish). In addition, we include 2K-4.5K utterances per speaker for AS and STL. For TSes, we include around 15K-20K utterances by default, for better voice quality. Each utterance on average contains 55-75 phonemes and lasts about 5 seconds. All audio data are recorded in studio-grade settings. The designs of our experiments, described in Section~\ref{sec:experiment_design}, ensure that all evaluations are controlled, i.e., any set of models compared in an evaluation have a single changing variable amongst them. Therefore, in Section~\ref{sec:experiment_design}, instead of repeating the description of the part of training dataset shared by all models, we will implicitly intend that the core dataset described above is always included, and emphasise on the changing variable in data composition when describing the models.

\subsection{Evaluation method}
\label{sec:eva_method}
We use a mixture of preference test and MUltiple Stimuli with Hidden Reference and Anchor (MUSHRA) test \cite{recommendation2001bs} as the means of subjective evaluation. As most evaluations we are conducting only involve comparing a pair of controlled models, we mainly rely on preference tests in which we ask test participants (native speakers of the LV being tested) to listen to samples from both models and indicate which audio sample they prefer overall. 
The preference test is essentially a binomial test. We use participants' votes to establish whether to accept our null hypothesis that one model has higher probability of being perceived as better than the other. If the preference test does not show any statistically significant preference for a particular model, we then conduct more granular MUSHRA tests in which we ask test participants to rate audio samples on a scale between 0 and 100, in terms of naturalness and accent. For the naturalness MUSHRA test, we use the audio recordings of the TS speaking their native language as the upper anchor. For accent evaluation, we use recordings from a native speaker of the synthesised language as upper anchor, and synthesis from a model that has been trained with monolingual data of the original speaker only as lower anchor. For the lower anchor, we map each phone of the language we want to synthesise with to a similar phone that exists in the TS's monolingual model, purposely producing accented speech. For both tests, we use the threshold of $5\%$ to indicate whether results are statistically significant. We synthesise 100 testing samples for all LVs being tested, and each test sample is evaluated 50/30 times for the preference/MUSHRA tests, respectively. 

\subsection{Experiment design}
\label{sec:experiment_design}
In this section, we describe the design of the experiments for understanding how various factors in data composition affect the TS's Polyglot synthesis in TLs. It should be noted that our experiments here are by no means exhaustive searches on all possible data compositions, which is infeasible in practice. Instead, we carefully design several controlled experiments to reveal the trend. The TSs, TLs and ASes for our experiments are selected to be representative, e.g. of a language family, while also considering data availability. All subjective evaluations conducted are controlled in terms of the number of training epochs.

\subsubsection{Experiment 1: Language family affiliation of AS}
\label{sec:exp1}
In this experiment, the goal is to understand how the language family affiliations of ASes affect the quality of the TSes speaking the TLs. To this end, we design 2 sets of experiments as follows.

\begin{table}[]
\caption{
\label{tab:exp1}
Models trained \& evaluated in Experiment 1.}
\centering
\begin{tabular}{l|l|l|l}
\toprule
\makecell{\textbf{Exp.}} & \makecell{\textbf{Model}} & \makecell{\textbf{TL}} & \makecell{\textbf{Difference in data composition}}                       \\ \midrule

\multirow{2}{*}{\begin{tabular}[c]{@{}l@{}} \makecell{1.1} \end{tabular}} &\makecell{M1}    & \multirow{2}{*}{\begin{tabular}[c]{@{}l@{}} \makecell{it-IT\\ de-DE} \end{tabular}} & \makecell{4 Romance language speakers}  \\ \cline{2-2} \cline{4-4} 

 & \makecell{M2}    &                                                                        & \makecell{4 Germanic language speakers} \\ \hline

\multirow{2}{*}{\begin{tabular}[c]{@{}l@{}} \makecell{1.2} \end{tabular}} & \makecell{M3}    & \multirow{2}{*}{\begin{tabular}[c]{@{}l@{}}\makecell{arb \\de-DE\\it-IT} \end{tabular}} & \makecell{Core dataset \\ and data from STL}  \\ \cline{2-2} \cline{4-4} 

& \makecell{M4}    &                                                                        &  \makecell{12 ASes from \\ multiple language families}
\\ \bottomrule
\end{tabular}

\vspace{.2cm}
Evaluation Results
\vspace{.1cm}

\begin{tabular}{l|l|l|l|l}
\toprule
\makecell{\textbf{LV}} & \makecell{\textbf{M1 vote}}  & \makecell{\textbf{No preference}} & \makecell{\textbf{M2 vote}} & \makecell{\textbf{p-value}} 
\\ \midrule

\makecell{de-DE}  & \makecell{34.65\%}  & \makecell{26.94\%} & \makecell{\textbf{38.41\%}} & \makecell{0.03\%}  
\\

\makecell{it-IT}  & \makecell{\textbf{50.41\%}}  & \makecell{21.78\%} & \makecell{27.82\%} & \makecell{0.00\%} 
\\ \midrule

\makecell{\textbf{LV}} & \makecell{\textbf{M3 vote}}  & \makecell{\textbf{No pref}} & \makecell{\textbf{M4 vote}}&  \makecell{\textbf{p-value}} 
\\ \hline 

\makecell{de-DE}  & \makecell{20.10\%}  & \makecell{14.28\%} & \makecell{\textbf{65.62\%}} & \makecell{0.00\%}  
\\ 

\makecell{it-IT}  & \makecell{26.82\%}  & \makecell{13.36\%} & \makecell{\textbf{59.82\%}} & \makecell{0.00\%}
\\

\makecell{arb}  & \makecell{37.45\%}  & \makecell{25.82\%} & \makecell{36.73\%} & \makecell{-}
\\ \bottomrule

\end{tabular}
\vspace{-.2cm}
\end{table}

\textbf{Experiment 1.1}: For the 2 models (M1 and M2) trained in this experiment, we select it-IT (a Romance language) and de-DE (a Germanic language) as TLs. The only difference between the 2 models is that in the training dataset of M1 there are 4 ASes from Romance languages (1 fr-FR, 1 es-ES, 1 pt-BR, 1 es-US) in addition to the core dataset, whereas M2 has 4 additional ASes from 4 Germanic languages (1 en-US, 1 nb-NO, 1 sv-SE, 1 is-IS). For evaluation, we expect to observe whether the addition of ASes from different language families affects the TLs belonging to different language families differently. 

\textbf{Experiment 1.2}: In Experiment 1.1 above, we only added ASes from the same language family in each model. We are also interested in knowing the effect of adding ASes belonging to multiple language families. Therefore, in Experiment 1.2, we add a basket of ASes from different language families, and evaluate how this would affect different TLs. Specifically, in model M3, the training data composition only includes the core dataset (described in Section~\ref{sec:baseline_dataset}) and data from STLs. By contrast, for training M4, we add 12 additional ASes in cy-GB, da-DA, is-IS, fr-FR, pl-PL, ro-RO, pt-BR, ru-RU, tr-TR, sv-SE, ko-KR, and nb-NO. There are 4 Germanic and 3 Romance language ASes amongst them, together with 5 ASes from other language families. For both M3 and M4, the TLs are selected to be arb, de-DE and it-IT. The addition of arb as TL in this experiment is meant to test whether adding ASes in distant language families from the TL has any bearing on the TL.  

Table~\ref{tab:exp1} summarises the experiment configurations. Note that all models evaluated in the same evaluation here have identical gender composition for speakers present in the datasets.

\subsubsection{Experiment 2: Gender mixture of AS}
\label{sec:exp2}
In this experiment, we investigate the influence of gender mixture of ASes on male and female TS's Polyglot synthesis, to answer questions pertaining to the gender mixture of the speakers in the training corpora. For this purpose, all models we train for this experiment have a male and a female TS. The TLs for this experiment are arb and it-IT. Then, we consider 3 groups of gender mixtures of ASes, i.e., all female, all male, and gender-balanced. The models trained for this experiment (M5, M6, and M7) are summarised in Table~\ref{tab:exp2}. Note that all models trained for Experiment 2 and 3 have identical language composition.

\begin{table}[]
\caption{
\label{tab:exp2}
Models trained \& evaluated in Experiment 2.}
\centering
\begin{tabular}{l|l|l|l}
\toprule
\makecell{\textbf{Model}} & \makecell{\textbf{TL}}    &   \makecell{\textbf{TS}} & \makecell{\textbf{Difference in data} \\\textbf{composition}} \\ \midrule

\makecell{M5}    & \multirow{3}{*}{\begin{tabular}[c]{@{}l@{}} \makecell{arb \\ it-IT} \end{tabular}}   & \multirow{3}{*}{\begin{tabular}[c]{@{}l@{}} \makecell{female \\ male} \end{tabular}}   &\makecell{8 female AS}  \\ \cline{1-1} \cline{4-4} 

\makecell{M6}    &   &   & \makecell{4 female AS + 4 male AS} \\ \cline{1-1} \cline{4-4} 

\makecell{M7}    &   &   & \makecell{8 male AS} 
\\ \bottomrule
\end{tabular}

\vspace{.2cm}
Evaluation Results
\vspace{.1cm}

\begin{tabular}{l|l|l|l|l}
\toprule
\multicolumn{5}{c}{\textbf{Female TS}}
\\ \midrule

\makecell{\textbf{LV}} & \makecell{\textbf{M5 vote}}  & \makecell{\textbf{No pref}} & \makecell{\textbf{M6 vote}}&  \makecell{\textbf{p-value}} 
\\ \hline

\makecell{it-IT}  & \makecell{\textbf{42.86\%}}  & \makecell{22.30\%} & \makecell{34.84\%} & \makecell{0.00\%} 
\\ 

\makecell{arb}  & \makecell{\textbf{39.70}\%}  & \makecell{25.90\%} & \makecell{34.40\%} & \makecell{0.00\%}  
\\ \midrule

\makecell{\textbf{LV}} & \makecell{\textbf{M6 vote}}  & \makecell{\textbf{No pref}} & \makecell{\textbf{M7 vote}} &  \makecell{\textbf{p-value}} 
\\ \hline

\makecell{it-IT}  & \makecell{32.49\%}  & \makecell{33.90\%} & \makecell{33.60\%} & \makecell{-} 
\\ 

\makecell{arb}  & \makecell{33.19\%}  & \makecell{33.54\%} & \makecell{33.28\%} & \makecell{-}  
\\ \midrule

\multicolumn{5}{c}{\textbf{Male TS}}
\\ \midrule
\makecell{\textbf{LV}} & \makecell{\textbf{M5 vote}}  & \makecell{\textbf{No pref}} & \makecell{\textbf{M6 vote}} &  \makecell{\textbf{p-value}} 
\\ \hline

\makecell{it-IT}  & \makecell{\textbf{43.96\%}}  & \makecell{24.92\%} & \makecell{31.12\%} & \makecell{0.00\%} 
\\ 

\makecell{arb}  & \makecell{\textbf{42.43\%}}  & \makecell{21.97\%} & \makecell{35.60\%} & \makecell{0.00\%}  
\\ \midrule

\makecell{\textbf{LV}} & \makecell{\textbf{M6 vote}}  & \makecell{\textbf{No pref}} & \makecell{\textbf{M7 vote}} &  \makecell{\textbf{p-value}} 
\\ \hline

\makecell{it-IT}  & \makecell{\textbf{37.71\%}}  & \makecell{31.38\%} & \makecell{30.91\%} & \makecell{0.00\%} 
\\ 

\makecell{arb}  & \makecell{32.87\%}  & \makecell{34.14\%} & \makecell{32.99\%} & \makecell{-}  
\\ \bottomrule

\end{tabular}
\vspace{-.2cm}
\end{table}

\subsubsection{Experiment 3: Gender mixture of STL}
\label{sec:exp3}
STLs differ from ASes in that in the end we expect the TSes to speak in the native LVs of STLs, but not of ASes. Therefore, we explore the gender mixture of STL separately in this experiment. Similar to Experiment 2 above, we train 3 models for Experiment 3, i.e., models with all female (M8), gender-balanced (M9), and all male (M10) STLs in the training dataset. Based on the availability of our voice data, we select de-DE, en-GB, and fr-CA as TLs. The comparison of models for this experiment can be found in Table~\ref{tab:exp3}. 

\begin{table}[]
\caption{
\label{tab:exp3}
Models trained \& evaluated in Experiment 3.}
\vspace{-.1cm}
\centering
\begin{tabular}{l|l|l|l}
\toprule
\makecell{\textbf{Model}} & \makecell{\textbf{TL}}    &   \makecell{\textbf{TS}}   & \makecell{\textbf{Difference in data} \\ \textbf{composition}}    
\\ \midrule

\makecell{M8}    & \multirow{3}{*}{\begin{tabular}[c]{@{}l@{}} \makecell{de-DE \\ en-GB \\ fr-CA} \end{tabular}}   & \multirow{3}{*}{\begin{tabular}[c]{@{}l@{}} \makecell{female \\ male} \end{tabular}}   &\makecell{8 female STL/TL}  
\\ \cline{1-1} \cline{4-4} 

\makecell{M9} &   &   & \makecell{4 female + 4 male STL/TL} 
 \\ \cline{1-1} \cline{4-4} 

\makecell{M10}    &   &   & \makecell{8 male STL/TL} 
\\ \bottomrule
\end{tabular}

\vspace{.2cm}
Evaluation Results
\vspace{.1cm}

\begin{tabular}{l|l|l|l|l}
\toprule
\multicolumn{5}{c}{\textbf{Female TS}}
\\ \midrule

\makecell{\textbf{LV}} & \makecell{\textbf{M8 vote}}  & \makecell{\textbf{No pref}} & \makecell{\textbf{M9 vote}} &  \makecell{\textbf{p-value}} 
\\ \hline

\makecell{de-DE}  & \makecell{\textbf{58.25\%}}  & \makecell{16.13\%} & \makecell{25.63\%} & \makecell{0.00\%} 
\\ 

\makecell{en-GB}  & \makecell{\textbf{52.85\%}}  & \makecell{10.38\%} & \makecell{36.77\%} & \makecell{0.00\%} 
\\ 

\makecell{fr-CA}  & \makecell{\textbf{52.29\%}}  & \makecell{18.14\%} & \makecell{29.58\%} & \makecell{0.00\%}  
\\ \midrule

\makecell{\textbf{LV}} & \makecell{\textbf{M9 vote}}  & \makecell{\textbf{No pref}} & \makecell{\textbf{M10 vote}} &  \makecell{\textbf{p-value}} 
\\ \hline

\makecell{de-DE}  & \makecell{40.45\%}  & \makecell{18.59\%} & \makecell{40.95\%} & \makecell{-} 
\\ 

\makecell{en-GB}  & \makecell{\textbf{47.48\%}}  & \makecell{11.12\%} & \makecell{41.40\%} & \makecell{0.00\%} 
\\

\makecell{fr-CA}  & \makecell{\textbf{38.57\%}}  & \makecell{27.87\%} & \makecell{33.56\%} & \makecell{0.00\%}  
\\ \midrule

\multicolumn{5}{c}{\textbf{Male TS}}
\\ \midrule
\makecell{\textbf{LV}} & \makecell{\textbf{M8 vote}}  & \makecell{\textbf{No pref}} & \makecell{\textbf{M9 vote}} &  \makecell{\textbf{p-value}} 
\\ \hline

\makecell{de-DE}  & \makecell{\textbf{43.28\%}}  & \makecell{23.89\%} & \makecell{32.83\%} & \makecell{0.00\%} 
\\ 

\makecell{en-GB}  & \makecell{\textbf{49.74\%}}  & \makecell{14.85\%} & \makecell{35.41\%} & \makecell{0.00\%} 
\\ 

\makecell{fr-CA}  & \makecell{\textbf{47.34\%}}  & \makecell{20.02\%} & \makecell{32.65\%} & \makecell{0.00\%}  
\\ \midrule

\makecell{\textbf{LV}} & \makecell{\textbf{M9 vote}}  & \makecell{\textbf{No pref}} & \makecell{\textbf{M10 vote}} &  \makecell{\textbf{p-value}} 
\\ \hline

\makecell{de-DE}  & \makecell{\textbf{52.66\%}}  & \makecell{18.52\%} & \makecell{28.82\%} & \makecell{0.00\%} 
\\ 

\makecell{en-GB}  & \makecell{\textbf{49.63\%}}  & \makecell{10.46\%} & \makecell{39.91\%} & \makecell{0.00\%} 
\\

\makecell{fr-CA}  & \makecell{\textbf{40.43\%}}  & \makecell{23.75\%} & \makecell{35.82\%} & \makecell{0.01\%}  
\\ \bottomrule

\end{tabular}
\vspace{-.1cm}
\end{table}

\subsubsection{Experiment 4: The number of STL}
\label{sec:exp4}
For a voice cloning-based multilingual TTS system, the quality of TS speaking the TL is directly affected by the amount of data from STL in the training dataset. Therefore, in this section we carry out experiments to examine the effect of the number of STL present in the training dataset. We aim to understand, for example, whether it is always better to have more STL in the training dataset. Intuitively, when there is a diverse training corpus in TL, we would expect the model to benefit from better disentanglement between language and speaker identity in the TL. In particular, for this experiment we use 4, 10, 16 STLs per TL for training 3 models M11, M12, and M13, respectively. The TLs here are en-GB, fr-CA, and de-DE. 

\begin{table}[h]
\caption{
\label{tab:exp4}
Models trained \& evaluated in Experiment 4.}
\centering
\vspace{-.1cm}
\begin{tabular}{l|l|l}
\toprule
\makecell{\textbf{Model}} & \makecell{\textbf{TL}} & \makecell{\textbf{Difference in data composition}}                       
\\ \midrule

\makecell{M11}  & \multirow{3}{*}{\begin{tabular}[c]{@{}l@{}} \makecell{en-GB \\ fr-CA \\ de-DE} \end{tabular}} & \makecell{4 STL/TL}  \\ \cline{1-1} \cline{3-3} 

\makecell{M12}    &     & \makecell{10 STL/TL} 
\\ \cline{1-1} \cline{3-3} 

\makecell{M13}    &     & \makecell{16 STL/TL}  
\\ \bottomrule

\end{tabular}

\vspace{.2cm}
Evaluation Results
\vspace{.1cm}

\begin{tabular}{l|l|l|l|l|l}
\toprule
\makecell{\textbf{Item}} & \makecell{\textbf{Model}} & \makecell{\textbf{en-GB}}    &   \makecell{\textbf{fr-CA}}  &   \makecell{\textbf{de-DE}} & \makecell{\textbf{Significance}}    \\ \midrule

\multirow{3}{*}{\begin{tabular}[c]{@{}l@{}} \makecell{Nat.} \end{tabular}} &\makecell{M11}    & \makecell{59.25}  & \makecell{67.72}  & \makecell{63.03} &\multirow{3}{*}{\begin{tabular}[c]{@{}l@{}} \makecell{all \\ difference \\ significant} \end{tabular}}  
\\ \cline{2-2} \cline{3-3}  \cline{4-4} \cline{5-5} 

& \makecell{M12}    &  \makecell{\textbf{60.25}} &  \makecell{\textbf{69.47}} &  \makecell{\textbf{64.38}} &  
\\ \cline{2-2} \cline{3-3}  \cline{4-4} \cline{5-5} 

& \makecell{M13}    &  \makecell{57.68} &  \makecell{65.33}  &  \makecell{61.68}&  
\\ \midrule

\multirow{3}{*}{\begin{tabular}[c]{@{}l@{}} \makecell{Acc.} \end{tabular}} &\makecell{M11}    & \makecell{63.12}  & \makecell{74.40}  & \makecell{74.38} &\multirow{3}{*}{\begin{tabular}[c]{@{}l@{}} \makecell{only diff. \\ M11\&M13 \\ insignificant } \end{tabular}}  
\\ \cline{2-2} \cline{3-3}  \cline{4-4} \cline{5-5} 

& \makecell{M12}    &  \makecell{\textbf{64.37}} &  \makecell{\textbf{75.95}}  &  \makecell{\textbf{76.28}}&  
\\ \cline{2-2} \cline{3-3}  \cline{4-4} \cline{5-5} 

& \makecell{M13}    &  \makecell{63.34} &  \makecell{74.03} &  \makecell{74.5} &  
\\ \bottomrule

\end{tabular}
\vspace{-.2cm}
\end{table}

\begin{comment}
\begin{table}[]
\caption{
\label{tab:result:exp1}
Evaluation results for Experiment 1.1 (M1 and M2 are compared for de-DE and it-IT) and Experiment 1.2 (M3 and M4 are compared for de-DE, it-IT, and arb).}
\centering
\begin{tabular}{l|l|l|l|l}
\toprule
\makecell{\textbf{LV}} & \makecell{\textbf{M1 vote}}  & \makecell{\textbf{No preference}} & \makecell{\textbf{M2 vote}} & \makecell{\textbf{p-value}} 
\\ \midrule

\makecell{de-DE}  & \makecell{34.65\%}  & \makecell{26.94\%} & \makecell{\textbf{38.41\%}} & \makecell{0.03\%}  
\\

\makecell{it-IT}  & \makecell{\textbf{50.41\%}}  & \makecell{21.78\%} & \makecell{27.82\%} & \makecell{0.00\%} 
\\ \midrule

\makecell{\textbf{LV}} & \makecell{\textbf{M3 vote}}  & \makecell{\textbf{No pref}} & \makecell{\textbf{M4 vote}}&  \makecell{\textbf{p-value}} 
\\ \hline 

\makecell{de-DE}  & \makecell{20.10\%}  & \makecell{14.28\%} & \makecell{\textbf{65.62\%}} & \makecell{0.00\%}  
\\ 

\makecell{it-IT}  & \makecell{26.82\%}  & \makecell{13.36\%} & \makecell{\textbf{59.82\%}} & \makecell{0.00\%}
\\

\makecell{arb}  & \makecell{37.45\%}  & \makecell{25.82\%} & \makecell{36.73\%} & \makecell{-}
\\ \bottomrule

\end{tabular}
\end{table}
\end{comment}

\section{Evaluation result and discussion}
\label{sec:result_analysis}

\subsection{Effects of language family affiliation of AS}
\label{sec:exp_1_result}
The evaluation results for Experiment 1 are summarised in Table~\ref{tab:exp1}, which clearly show that adding ASes speaking languages in the same language family as the TL improves TS's polyglot synthesis in that TL. Furthermore, evaluation results from comparing M3 and M4 show that adding together ASes from several language families benefits the TLs which share the language family affiliation of ASes (see Section~\ref{sec:exp1} for detailed composition). The result also suggests that the extent of improvement for a TL is related to the amount of data from the ASes in the same language family as that of the TL, as de-DE receives higher percentage of votes than it-IT (the number of Germanic utterances is $80\%$ more than that of Romance utterances). For the other TL, arb, M3 and M4 are equally preferred. This can be explained by the absence of additional languages that are close to Arabic from M3's to M4's data compositions. We conclude that adding AS whose language is close to the language of TL can help improve the TL while adding ASes far from the TL makes little difference for the TL.  \looseness=-1

\subsection{Effects of gender mixture of AS and STL}
\label{sec:exp_2_3_result}
Tables~\ref{tab:exp2} and \ref{tab:exp3} summarise the evaluation results for Experiments 2 and 3, which test the influence of gender composition of AS and STL, respectively. The differences in AS/STL data composition can be also be found there. As the results from both experiments suggest the same trend in terms of preferred AS/STL gender mixture, we report them together in this section. A brief summary for this is that \textit{all female} $>$ \textit{gender-balanced} $\geq$ \textit{all male} for AS and STL data composition. This is true for both female and male TSes. In addition, the difference between gender-balanced and all male in AS/STL is more pronounced for male TS, as there are more cases where there is statistically significant preference for gender-balanced AS/STL over all male AS/STL. Informal subjective evaluations found that for male TS, models trained with more female data tend to reduce the creakiness of TS's voice speaking the TLs. The benefits of having more female AS/STL is less noticeable for female TS's, but we nonetheless observed improved articulation clarity in some cases. This result is surprising as we did not observe any speaker identity issues, i.e., the TSes still keep their speaker identities even when all training data in the TLs is from STLs of the opposite gender. This is a strong data point which indicates that by including the core dataset described in Section~\ref{sec:baseline_dataset}, the model achieved good speaker-language disentanglement. \looseness=-1

\subsection{Effects of the amount of STL data}
\label{sec:exp_4_result}
In Table~\ref{tab:exp4}, we present evaluation results from MUSHRA tests for comparing M11, M12, and M13. 
As mentioned in Section~\ref{sec:eva_method}, here we use MUSHRA tests for evaluating naturalness and accent separately, as preference tests failed to produce statistically significant distinctions between the models. From the MUSHRA scores in the table, we can see that having 10 STL produces the best scores for both naturalness and accent, for all 3 LVs tested. For naturalness, there is \textit{10 STL} $>$ \textit{4 STL} $>$ \textit{16 STL}; whereas we have \textit{10 STL} $>$ \textit{4 STL} $=$ \textit{16 STL} for accent. This result refutes our original hypothesis that having more STLes always improves the model's TL performance. Instead, we observe that there is a sweet spot, and more data is not always better. This might be explained by the fact that the training dataset becomes more unbalanced towards the TLs, as we keep adding STL data, which in turn hinders the model’s ability to
generalise and disentangle different LVs.

\section{Conclusions}
\label{sec:conclusion}
In this paper, we studied how the data composition of training corpora affects multilingual NTTS synthesis in terms of the language family affiliation, gender balance and the number of speakers present. Our most surprising result is that including female speakers is better than having a balanced mixture of genders. Also, supporting speakers whose native LVs are closer to the target language are beneficial. Moreover, we conclude that adding more speakers in the target LV improves the model only up to a point, after which adding more incurs regressions. While these results are useful guidelines for building training corpora for multilingual NTTS systems, we are not yet able to fully explain some of them, which will be investigated as future work.

\bibliographystyle{IEEEtran}

\end{document}